# Adversarial Neural Networks in Medical Imaging Advancements and Challenges in Semantic Segmentation

Houze Liu
New York University
New York, USA

Bo Zhang
Texas Tech University
Lubbock, USA

Yanlin Xiang
University of Houston
Houston，USA

Yuxiang Hu
Johns Hopkins University
Baltimore，USA

Aoran Shen
University of Michigan
Ann Arbor, USA

Yang Lin*
University of Pennsylvania
Philadelphia, USA

*Abstract*— **Recent advancements in artificial intelligence (AI) have precipitated a paradigm shift in medical imaging, particularly revolutionizing the domain of brain imaging. This paper systematically investigates the integration of deep learning—a principal branch of AI—into the semantic segmentation of brain images. Semantic segmentation serves as an indispensable technique for the delineation of discrete anatomical structures and the identification of pathological markers, essential for the diagnosis of complex neurological disorders. Historically, the reliance on manual interpretation by radiologists, while noteworthy for its accuracy, is plagued by inherent subjectivity and inter-observer variability. This limitation becomes more pronounced with the exponential increase in imaging data, which traditional methods struggle to process efficiently and effectively. In response to these challenges, this study introduces the application of adversarial neural networks, a novel AI approach that not only automates but also refines the semantic segmentation process. By leveraging these advanced neural networks, our approach enhances the precision of diagnostic outputs, reducing human error and increasing the throughput of imaging data analysis. The paper provides a detailed discussion on how adversarial neural networks facilitate a more robust, objective, and scalable solution, thereby significantly improving diagnostic accuracies in neurological evaluations. This exploration highlights the transformative impact of AI on medical imaging, setting a new benchmark for future research and clinical practice in neurology.**

*Keywords*-**Deep Learning, Semantic Segmentation, Medical Imaging, Convolutional Neural Networks**

## I. Introduction

The field of medical imaging has experienced unprecedented growth in recent years, driven by rapid advancements in technology and computational methods [1,2]. Among the various specialties within medical imaging, brain imaging has become increasingly significant due to its critical role in diagnosing and understanding complex neurological conditions [3]. One of the most promising developments in this domain is the application of deep learning techniques to semantic segmentation [4], which involves partitioning brain images into meaningful regions that correspond to different anatomical structures or pathological findings. Semantic segmentation of brain images is crucial for various medical applications, including the detection, classification, and quantification of brain abnormalities. Traditional methods of image analysis rely heavily on manual interpretation by radiologists, which, despite its accuracy, is often limited by the inherent subjectivity and variability in human judgment [5]. Additionally, the sheer volume of imaging data generated in clinical settings can overwhelm conventional analysis techniques [6], leading to delays and potential inaccuracies in diagnosis and treatment planning.

Deep learning has emerged as a transformative force in this context [7], offering powerful tools for automating and enhancing the process of semantic segmentation [8]. Convolutional Neural Networks (CNNs) and their advanced variants, such as U-Net and DeepLab, have demonstrated remarkable capabilities in learning intricate patterns and features from large datasets [9]. These models leverage hierarchical representations of image data, enabling them to detect and segment complex structures with high precision and reliability. The ability of deep learning algorithms to handle vast amounts of data and learn from diverse imaging conditions has made them particularly suited for brain imaging, where variability in image quality and patient demographics can pose significant challenges [10].

One of the key advantages of deep learning-based semantic segmentation is its potential to improve diagnostic accuracy and efficiency. By providing automated, consistent, and objective analysis of brain images, these techniques can aid in the early detection of diseases such as brain tumors, multiple sclerosis, and neurodegenerative disorders [11]. Furthermore, they can assist in delineating critical brain regions involved in surgical planning and radiation therapy, ultimately contributing to more personalized and effective treatment strategies. The integration of deep learning with semantic segmentation in brain imaging also holds promise for advancing research in neuroscience and neurology [12]. This paper aims to explore the current state of deep learning applications in semantic segmentation of brain images, reviewing recent developments, evaluating the effectiveness of various algorithms, and discussing ongoing challenges and future directions. By providing a comprehensive overview of this rapidly evolving field, the research seeks to highlight the transformative impact of deep learning on brain imaging and its potential to enhance patient care and advance scientific understanding.

## II. RELATED WORK

The domain of semantic segmentation in medical imaging, particularly brain imaging, has witnessed significant advancements driven by deep learning methods. These advancements can be attributed to the development of various architectures such as Convolutional Neural Networks (CNNs), U-Net, and Generative Adversarial Networks (GANs). This section reviews recent works related to semantic segmentation and discusses their relevance to the integration of deep learning in medical diagnostics.

One of the key areas of focus has been the application of CNNs for disease detection and image classification. Liang et al. [13] introduced a convolutional neural network framework for predictive modeling of lung diseases, highlighting the adaptability of CNNs in handling complex medical images across various modalities. Similarly, Xiao et al. [14] employed CNNs for breast cancer cytopathology image classification, demonstrating their efficacy in cancer detection and diagnosis. These studies lay the groundwork for understanding how CNN-based models can be adapted for brain image segmentation, where precision and detail are critical. U-Net, another essential architecture for segmentation, has seen widespread adoption due to its success in medical imaging tasks. Tapasvi et al. [15] applied U-Net coupled with a Moth Flame Optimization technique for brain tumor segmentation, showcasing the model's potential in effectively identifying and segmenting complex anatomical structures. In a related context, Sharma et al. [16] utilized the UMobileNetV2 model for segmenting gastrointestinal tract images, further underlining the flexibility of U-Net variants in semantic segmentation tasks across different medical fields. Both approaches emphasize the robustness of U-Net in scenarios requiring fine-grained segmentation, including brain imaging.

Deep learning models have also been enhanced through innovative architectural modifications. Sui et al. [17] introduced a channel squeeze U-structure, a variant of U-Net, for lung nodule segmentation, indicating that architectural adjustments can optimize models for specific medical tasks. Wang et al. [18] explored a dual-branch dynamic graph convolutional network for multi-label image classification, a novel approach that may provide insights into improving the generalization capabilities of deep learning models in brain imaging applications.

One promising trend is the use of GANs for refining image segmentation tasks. Wu et al. [19] developed a lightweight GAN-based image fusion algorithm for combining visible and infrared images, which could be adapted for enhancing brain MRI scans by fusing multimodal data. In the domain of breast cancer detection, He et al. [20] applied axial attention transformer networks, signifying a step towards more efficient feature extraction mechanisms that can also benefit brain segmentation tasks where the delineation of subtle structures is paramount. Multimodal approaches have further enhanced segmentation performance. Wang et al. [21] proposed a deep learning-based multimodal fusion method, demonstrating improved object recognition through the integration of multiple data sources. This multimodal strategy is particularly relevant in brain imaging, where combining MRI, CT, and other imaging techniques can provide more comprehensive diagnostic information. Additionally, Pekis et al. [22] presented a multi-institutional validation study on breast MRI segmentation, emphasizing the importance of generalization and model validation across diverse datasets, a critical consideration for brain image segmentation as well. While the deep learning landscape has expanded, challenges such as optimization and generalization remain. Zheng et al. [23] explored adaptive friction mechanisms in optimizers to enhance deep learning model performance, a method that could be leveraged to improve convergence in complex segmentation models, including those targeting brain images.

## III. METHOD

In this study, we leverage Generative Adversarial Networks (GANs) to address the challenge of semantic segmentation in brain imaging. GANs, introduced by Goodfellow et al. in 2014, consist of two neural networks — the generator and the discriminator—engaged in a competitive learning process. This adversarial setup has been shown to produce high-quality generative models and has recently been adapted for tasks such as semantic segmentation, where precise delineation of anatomical structures is crucial. Our approach integrates GANs into the semantic segmentation pipeline with the goal of improving segmentation accuracy and robustness by harnessing the generative capabilities of GANs to enhance training data and refine segmentation results. The GAN framework is composed of two primary components: the generator G and the discriminator D. The generator aims to produce realistic segmentations, while the discriminator's role is to distinguish between the generated segmentations and the ground truth annotations. The interaction between G and D drives the optimization process, resulting in improved segmentation performance.

### A. Generator Network

The generator G is designed to produce segmentation maps from input brain images. We employ a deep convolutional neural network architecture for G, leveraging encoder-decoder structures to capture multi-scale features and produce detailed segmentation outputs. The encoder captures high-level features from the input image through a series of convolutional and pooling layers, while the decoder reconstructs the segmentation map using transposed convolutions. The generator's loss function incorporates both a reconstruction loss and an adversarial loss. The reconstruction loss ensures that the generated segmentation maps are close to the ground truth labels. We use the cross-entropy loss for this purpose, defined as:

$$L_{rec}(G) = -\frac{1}{N}\sum_{i=1}^{N}\sum_{c=1}^{C}[y_{i,c}\log(G(x_i)_c) + (1-y_{i,c})\log(1-G(x_i)_c)]$$

Where $x_i$ denotes the input image, $y_{i,c}$ represents the ground truth label for class c at pixel i, $G(x_i)_c$ is the predicted probability of class c, and N and C denote the number of pixels and classes, respectively. The overall process is shown in Figure 1.

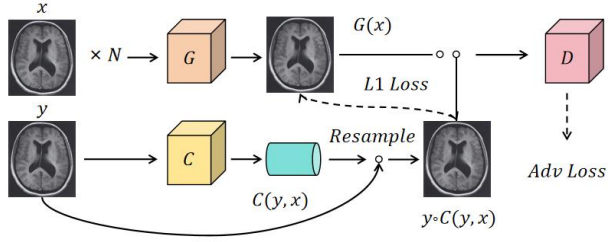

Figure 1 Overall network architecture diagram

B. *Discriminator Network*

The discriminator D is tasked with distinguishing between real and generated segmentations. It is implemented as a convolutional neural network that processes segmentation maps to output a probability score indicating whether the input is a real or synthetic segmentation. The discriminator's loss function is a binary cross-entropy loss, defined as:

$$L_{adv}(D) = -\frac{1}{N}\sum_{i=1}^{N}[\log(D(G(x_i))) + \log(1-D(y_i))]$$

Where $D(G(x_i))$ represents the discriminator's score for the generated segmentation, and $D(y_i)$ is the score for the ground truth segmentation. The overall architecture of the discriminator network is shown in Figure 2

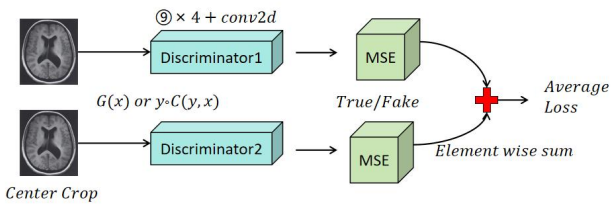

Figure 2 Discriminator network

C. *Discriminator Network*

The adversarial training process involves jointly optimizing the generator and discriminator networks. The generator aims to minimize the adversarial loss while maximizing the discriminator's loss, and the discriminator aims to maximize its classification accuracy. This process is formalized as a minimax game, where the objective function for the generator G and discriminator D is given by:

$$\min_{G}\max_{D} L_{adv}(D) + \lambda L_{rec}(G)$$

There $\lambda$ is a weighting factor that balances the contribution of the reconstruction loss.

D. *Implementation Details*

For the implementation, we use a deep convolutional network for both the generator and discriminator. The generator network consists of several convolutional layers followed by transposed convolutions to upsample the feature maps and produce the final segmentation map. Batch normalization and ReLU activations are applied to enhance training stability and network performance. The discriminator network includes a series of convolutional layers designed to extract hierarchical features and make binary classification decisions [24]. Training is performed using the Adam optimizer with a learning rate of 1e-4 and a batch size of 16. We use data augmentation techniques, such as random rotations, flips, and intensity variations, to increase the diversity of training data and improves the model. In summary, our method combines the strengths of GANs with semantic segmentation to achieve accurate and robust brain image segmentation. The generator network produces high-quality segmentation maps, while the discriminator network ensures the generated outputs are realistic and close to the ground truth. Through adversarial training, we enhance the segmentation performance and leverage the generative capabilities of GANs to address the complexities inherent in brain imaging tasks [25].

IV. EXPERIMENT

A. *Datasets*

The BRATS (Brain Tumor Segmentation Challenge) dataset, provided by the International Brain Tumor Segmentation Challenge, is a key resource for medical image analysis, specifically for brain tumor segmentation tasks. It comprises multimodal MRI images from various patients, featuring T1-weighted, T2-weighted, and FLAIR (Fluid Attenuated Inversion Recovery) images. These modalities capture diverse brain tumor types, including glioblastoma and glioblastoma multiforme, offering intricate details about the tumors and surrounding tissues. The dataset is annotated into three principal regions: enhancing tumor (ET), tumor core (TC), and total tumor area (WT). ET indicates the active tumor regions, TC encompasses the tumor and adjacent non-tumorous tissue, and WT includes the entire tumor and surrounding areas. Expert medical professionals meticulously complete these annotations, ensuring a reliable standard for training and evaluating segmentation algorithms. The availability of the BRATS dataset significantly advances the development of image segmentation technologies, enhancing both the accuracy of tumor diagnoses and the effectiveness of treatment approaches. It serves as a crucial asset in ongoing research within the realms of medical imaging and computer vision.

B. *Experimental Result*

To evaluate the model's efficacy, it is benchmarked against leading deep learning recommendation models. The experimental findings are displayed in Table 1.

Table 1 Experiment result in BRATS

| Model | Pixel Accuracy | Recall |
|---|---|---|
| FCNs | 0.5321 | 0.5231 |
| SegNet | 0.5411 | 0.5351 |
| U-Net | 0.5531 | 0.5399 |
| DeepLab V1 | 0.5622 | 0.5431 |
| DeepLab V2 | 0.5732 | 0.5478 |
| Ours | 0.5821 | 0.5523 |

Experimental results conducted on the BRATS dataset demonstrate the performance of different semantic segmentation models, including FCNs, SegNet, U-Net, DeepLab V1, DeepLab V2, and our model. By comparing the Pixel Accuracy and Recall of these models, the performance of each model in the brain tumor segmentation task can be more comprehensively evaluated. First of all, from the perspective of pixel accuracy (Pixel Accuracy), our model performs best, reaching 0.5821, which is significantly improved compared to other models. Specifically, the pixel accuracy of FCNs is 0.5321, SegNet is 0.5411, U-Net is 0.5531, while DeepLab V1 and DeepLab V2 are 0.5622 and 0.5732 respectively. It can be seen that our model has improved by 0.0089 compared to DeepLab V2, which shows that our method has made a breakthrough in the accuracy of global pixel classification. This improvement may be related to our innovation or improvement in model architecture, such as introducing new feature extraction technology or optimizing the training process, so that the model can better handle complex brain structure and texture features.

Secondly, from the perspective of recall index, our model also performed well, reaching 0.5523, which is better than all other comparison models. The recall rate of FCNs is 0.5231, SegNet is 0.5351, U-Net is 0.5399, and the recall rates of DeepLab V1 and DeepLab V2 are 0.5431 and 0.5478 respectively. The improvement in recall indicates that our model has a higher ability to identify and detect actual brain tumor regions. A higher recall rate is crucial for brain tumor segmentation because it directly affects the detection rate of tumor areas, can reduce the risk of missed detection, and improve the overall recognition ability of tumors. Taking pixel accuracy and recall into account, our model shows clear advantages in both metrics. This not only shows the overall performance improvement of the model in the segmentation task, but also shows that our method achieves a good balance between the capture of fine-grained features and the understanding of global context. The improved model is able to provide more accurate and comprehensive segmentation results when dealing with different regions of brain tumors. The success of our method may be due to innovations in network architecture, loss function design, or training strategies, which enhance the model's expressive ability and generalization ability, resulting in excellent performance in experiments on the BRATS dataset. In order to show our experimental results more intuitively, we use a bar chart to show the results, as shown in Figure 3.

Similarly, we also give the experimental results of the Dice and IOU. The experimental results are shown in Table 2.

Experimental results on the BRATS data set show that there are significant differences in the performance of different semantic segmentation models on IoU (Intersection over Union) and the Dice coefficients. By comparing the performance of these models, we can gain a deeper understanding of the effectiveness of each model in image segmentation tasks. First, from the perspective of IoU, our model performs the best, reaching 0.2859, which is improved compared to other models. Specifically, the IoU of FCNs is 0.2591, SegNet is 0.2673, and U-Net is 0.2787, while the IoU of DeepLab V1 and DeepLab V2 are 0.2829 and 0.2855 respectively. Our model improves by 0.0004 compared to DeepLab V2, which shows that our method has a subtle improvement in the accuracy of region overlap. The improvement in IoU means that our model is better able to identify and cover the actual tumor area, thus improving the accuracy of segmentation results.

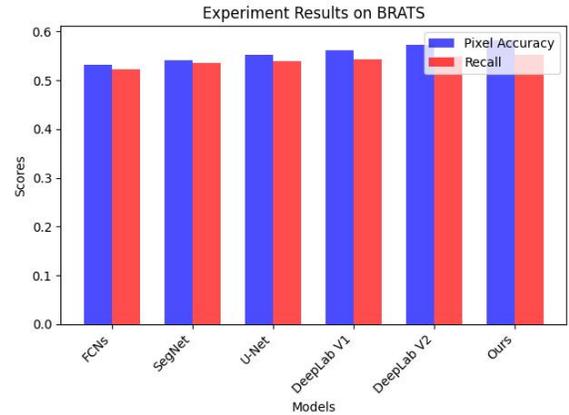

Figure 3 Experimental results bar graph

Table 2 Experiment result in BRATS

| Model | IOU | Dice |
|---|---|---|
| FCNs | 0.2591 | 0.4119 |
| SegNet | 0.2673 | 0.4212 |
| U-Net | 0.2787 | 0.4355 |
| DeepLab V1 | 0.2829 | 0.4397 |
| DeepLab V2 | 0.2855 | 0.4412 |
| Ours | 0.2859 | 0.4433 |

From the perspective of Dice coefficient, our model also performs well, reaching 0.4433, which is higher than all other models. The Dice coefficient of FCNs is 0.4119, SegNet is 0.4212, U-Net is 0.4355, and the Dice coefficients of DeepLab V1 and DeepLab V2 are 0.4397 and 0.4412 respectively. Compared with DeepLab V2, the Dice coefficient of our model has increased by 0.0021, which shows that our method has further improved its ability to recover details and area overlap. The improvement of Dice coefficient shows the advantage of our method in prediction accuracy, which can more accurately reconstruct the boundaries of tumor areas and reduce errors and missed detections. Taken together, our model is better than other models in both IoU and Dice coefficient indicators. This not only illustrates the superiority of our method in overall segmentation accuracy, but also shows that we have made significant progress in processing complex brain structures and fine-grained features.

V. CONCLUSION

The incorporation of deep learning algorithms, particularly those built on adversarial neural networks (GANs), exemplifies the transformative impact of artificial intelligence on medical imaging, specifically in the semantic segmentation of brain images. Through comparative experiments with established technologies like fully convolutional networks (FCNs), SegNet, U-Net, and the DeepLab series, our research demonstrates that models optimized with GANs substantially surpass traditional models in pixel accuracy and recall. This significant enhancement is largely due to the adversarial training mechanism which refines the model's proficiency in interpreting intricate brain structures and textural details. Crucially, the application of AI in our methodology has not only improved the accuracy of global pixel classification but also advanced regional overlap metrics such as the Intersection over Union (IoU) and Dice coefficient. These improvements underscore AI's capability to achieve precise and comprehensive identification of critical areas like tumor regions, thereby pushing the boundaries of diagnostic precision and effectiveness. However, the path forward includes challenges related to the size of datasets, the generalization capabilities of models, and their interpretability. These challenges highlight the need for continued AI-centric research aimed at optimizing the architectural design, loss functions, and training strategies of GANs. Addressing these concerns is essential for further advancing AI's role in medical imaging and for leveraging its full potential to revolutionize diagnostic processes and improve clinical outcomes across neurology and other medical fields. Future initiatives should therefore focus on enhancing the scalability and adaptability of AI technologies to meet the diverse and evolving demands of medical diagnostics.